\documentclass[pre,aps,showpacs,showkeys]{revtex4}

\usepackage{amsfonts}
\usepackage{amssymb}
\usepackage{amsmath}
\usepackage{latexsym}

\usepackage[dvips]{graphicx}

\newcommand{\sikib}{\begin{eqnarray}}
\newcommand{\sikie}{\end{eqnarray}}
\newcommand{\sikibnon}{\begin{eqnarray*}}
\newcommand{\sikienon}{\end{eqnarray*}}

\newcommand{\pd}[2]{\dfrac{\partial #1}{\partial #2}}
\newcommand{\pdd}[2]{\dfrac{\partial^2 #1}{\partial #2^2}}

\newcommand{\gin}[1]{g_{in}(\vec{#1})}
\newcommand{\gout}[1]{g_{out}(\vec{#1})}

\newcommand{\erad}[1]{e_{rad}(\vec{#1})}
\newcommand{\srad}[1]{s_{rad}(\vec{#1})}
\newcommand{\frad}[1]{f_{rad}(\vec{#1})}
\newcommand{\Trad}[1]{T_{rad}(\vec{#1})}
\newcommand{\psirad}[1]{\psi(\vec{#1})}

\newcommand{\pt}{{\bf P}}
\newcommand{\Prad}[1]{P_{rad}(\vec{#1})}

\begin{document}

\title{Two-temperature Steady State Thermodynamics for a Radiation Field}

\author{Hiromi Saida}
\email{saida@daido-it.ac.jp}

\affiliation{Department of Physics, Daido Institute of Technology\\
             Takiharu-cho 10-3, Minami-ku, Nagoya 457-8530, Japan}


\begin{abstract}
A candidate for a consistent steady state thermodynamics is constructed for a radiation field in vacuum sandwiched by two black bodies of different temperatures. Because of the collisionless nature of photons, a steady state of a radiation field is completely determined by the temperatures of the two black bodies. Then the zeroth, first, second and third laws can be extended to steady states, where the idea of local steady states plays an important role for the system whose geometrical shape is anisotropic and inhomogeneous. The thermodynamic formalism presented in this paper does not include an energy flux as a state variable. This is consistent with the notable conclusion by {\it C. Essex, Adv. Thermodyn. 3 (1990) 435; Planet. Space. Sci. 32 (1984) 1035} that, contrary to the success in the irreversible thermodynamics for dissipative systems, a nonequilibrium radiation field does not obey the bilinear formalism of the entropy production rate using an energy flux and its conjugate force. Though the formalism given in this paper may be unique to a radiation field, a nonequilibrium order parameter of steady states of a radiation field is explicitly defined. This order parameter denotes that the geometrical shape of the system determines how a steady state is far from an equilibrium. The higher the geometrical symmetry, the more distant the steady state.
\end{abstract}



\maketitle
\section{Introduction}\label{sec-intro}

The traditional treatment of the radiative energy transfer \cite{ref-stellar} has been applied to a mixture of a radiation field with a matter like a dense gas or other medium. In such a traditional treatment, the successive absorption and emission of photons by components of the medium matter makes it possible to consider that the photons are as if in local equilibrium states whose temperatures equal those of local equilibrium states of the medium matter. However, because photons are collisionless, it is impossible to apply the idea of local equilibrium to a radiation field {\it in vacuum}. 

By the way, when one deals with nonequilibrium states for a matter (without including a radiation field), it is usual to place an interest on the heat flux. Indeed for example, the classical and extended irreversible thermodynamics \cite{ref-cit} \cite{ref-eit} make use of the bilinear formalism of the entropy production rate using the heat flux and its conjugate thermodynamic force to obtain many developments in describing nonequilibrium phenomena. Then consider the case that a matter is put into a space where nothing but a nonequilibrium radiation field exists. 
This system possesses an interaction of the matter with the outside radiation field by the absorption and emission of photons at the surface of the matter. Note that, concerning a thermodynamic treatment of this system, it has already been revealed in references \cite{ref-essex} \cite{ref-wildt} that, because of the collisionless nature of photons {\it in vacuum}, the entropy production rate for the whole system composed of the matter and the radiation field can not be expressed in the bilinear form. And it is also stated in \cite{ref-essex} \cite{ref-wildt} that, even if the bilinear formalism was applied to this system, disappearing radiative energy flux (or thermodynamic force of the radiation field) would not always denote disappearing thermodynamic force (or radiative energy flux). This means that the energy flux does not work as a consistent state variable for a system including a nonequilibrium radiation field in vacuum. Then two conclusions follow, that the traditional treatment of the radiative transfer mentioned above is applicable only to a mixture of a radiation field with a matter which is dense enough to ignore the vacuum region among components of the matter, and that a distinct research from that of ordinary dissipative matter systems is required to understand a nonequilibrium radiation field {\it in vacuum}. However, a consistent thermodynamic formulation for a system including a nonequilibrium radiation field in vacuum has not been accomplished, and a consistent nonequilibrium order parameter for a radiation field has not been obtained so far. 

This paper shows a candidate for a consistent steady state thermodynamics for a radiation field, and obtains the consistent nonequilibrium order parameters of the steady states for a radiation field.
\footnote{There is a preprint {\it cond-mat/0310685} by H.S. which discusses a candidate for a free energy of a radiation field in a steady state. It is now noticed that the result of that preprint is wrong and can not formulate a consistent thermodynamic formalism for a radiation field in the cavity case. The resultant form obtained in this paper should be a right one of free energy.} 
We consider a radiation field {\it in vacuum} sandwiched by two {\it black bodies} of different temperatures. Keeping each temperature constant, the radiation field is in a nonequilibrium steady state with a stationary energy flow given by the Stefan-Boltzmann law. The settings treated in this paper are illustrated in figure \ref{pic.1}. Here it should be emphasised that, as pointed out at \S63 in reference \cite{ref-l.l}, the collisionless nature of photons denotes that a nonequilibrium radiation field can never relax to any equilibrium state unless there exists any interaction with other matter. Therefore the existence of the two black bodies in our settings is essential in formulating the steady state thermodynamics for a radiation field. Especially in extending the second law to steady states, it is necessary to consider a relaxation process of the whole system composed of the radiation field and the two black bodies. Given a well defined steady state entropy of the radiation field, the total entropy of the whole system have to be monotone increasing during the relaxation process of the whole system. 

Before proceeding to the construction of the steady state thermodynamics for a radiation field, it is helpful for readers the steady states of ordinary dissipative systems and the fluctuation of our radiation system are mentioned. As mentioned above in second paragraph, the thermodynamic formulation presented in this paper is unique to a radiation field. Therefore, when one attempts to construct a steady state thermodynamics for dissipative systems (without including a radiation field), a consideration different from ours should be made. Indeed, the reference \cite{ref-s.t} suggests a consistent macroscopic formulation of steady state thermodynamics for some dissipative systems like heat conduction, shear flow, electrical conduction and so on. In their formalism, the heat flux plays an essential role as a consistent state variable contrary to our radiation system, although it is also reported \cite{ref-h.n} that their steady state thermodynamics is not omnipotent and is applicable to some restricted class of dissipative systems. Further in reference \cite{ref-k.h-1}, the velocity distribution function of the steady state Boltzmann equation for a hard core molecule gas has already been explicitly derived up to second order in the density and temperature gradient. Their distribution function does also depend on the heat flux. Here we should note that, although it may appear strange that our formalism for a radiation field does not include an energy flux as a state variable, it seems that the heat flux plays the role of a consistent state variable for dissipative systems. The absence of an energy flux from consistent state variables should be understood as a unique property to a radiation field.

We next turn to the fluctuation of our radiation system. Because photons are collisionless, the distribution functions of photons in our settings are constructed by simply modifying an equilibrium Planckian distribution as shown in following sections. Consequently, the steady state of the radiation field is completely determined by the temperatures of the two black bodies. In addition, the collisionless nature shows that no fluctuation arises from the radiation field itself. The only possibility of a fluctuation arises from the absorption and emission processes of photons at the surfaces of the two black bodies. If the energy flux in the radiation field becomes strong enough and if the time scales of thermalisation of two black bodies are long enough, then the black bodies may be affected as to go into nonequilibrium states. In such a case, nonequilibrium temperatures of the bodies are unknown in general, and the distribution function of photons should be determined according to nonequilibrium statistical properties of the bodies. Concerning such a totally nonequilibrium case, there have already been some reports \cite{ref-d.f} \cite{ref-f.j.l} \cite{ref-fort} trying to determine the distribution function for a radiation field emitted by a nonequilibrium matter. They consider a radiative energy transfer inside a dense matter, and apply the information theory to the whole system composed of a radiation field and a medium matter.
\footnote{Recall the third paragraph in section \ref{sec-intro} that an isolated radiation field can never relax to any equilibrium state. Hence, in applying the information theory to a nonequilibrium radiation field, the reports \cite{ref-d.f} \cite{ref-f.j.l} \cite{ref-fort} considered the total entropy of the whole system composed of the radiation field and the medium matter.} 
However, after those reports were published, it has been revealed in reference \cite{ref-k.h-2} that, at least for a matter whose components are colliding and interacting with each other, the distribution function for a steady state of the matter derived with the information theory does not qualitatively agree with that derived with a steady state Boltzmann equation. Further it has also been concluded in \cite{ref-k.h-2} that the nonequilibrium temperature determined with the information theory has no physical meaning. Therefore, because their distribution function of the radiation field depends on the nonequilibrium temperature of the medium matter derived with the information theory, the reliability of the distribution function of the radiation field may not be given in those reports \cite{ref-d.f} \cite{ref-f.j.l} \cite{ref-fort}. Hence there is no confirmed form of the distribution function of a radiation field emitted by a nonequilibrium matter. In this paper, as a first step to construct a consistent macroscopic theory for a radiation field, we simply assume that each black body is always in an equilibrium state and no fluctuation arises. 

The contents of this paper are organised as follows. In section \ref{sec-require}, our basic requirements are explained. Then the steady state thermodynamics is established for the cylinder case (left one in figure \ref{pic.1}) in section \ref{sec-cylinder}, where a nonequilibrium order parameter of steady states is defined. Section \ref{sec-cavity} is devoted to the extension of the results obtained in section \ref{sec-cylinder} to the cavity case (right one in figure \ref{pic.1}), where the idea of local steady states plays an essential role. Summary and discussions are in section \ref{sec-sd}, where special properties of a radiation field, the distance of steady states from equilibrium states, future improvements and possible applications are discussed. 

Throughout this paper, $c$ is the speed of light, $\hbar = h/2 \pi$ is the Planck constant, $\sigma = \pi^2/60 \hbar^3 c^2$ is the Stefan-Boltzmann constant and the Boltzmann constant unity $k_B = 1$ is set.

\section{Basic requirements and conditions}\label{sec-require}

\begin{figure}[t]
 \begin{center}
 \includegraphics[height=25mm]{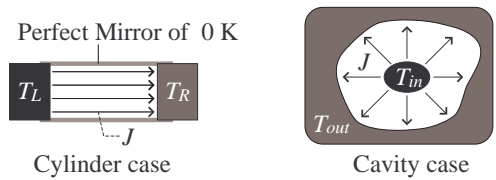}
 \end{center}
\caption{A steady state for a radiation field. $J$ is a stationary energy flow.}
\label{pic.1}
\end{figure}

We begin with the following basic requirements for constructing the steady state thermodynamics (SST) for a radiation field:
\begin{description}
 \item[(R1)] The state variables of equilibrium thermodynamics are extended to steady states. And at equilibrium limits, those extended variables take on the same value as the equilibrium ones.
 \item[(R2)] There exists a state variable which describes a nonequilibrium order of steady states. And that variable disappears at equilibrium limits. 
\end{description}
These requirements denote the zeroth law of the SST, the existence of steady states. A steady state of a radiation field is realised in the settings shown in figure \ref{pic.1}. In the cylinder case, the left and right lids are black bodies and in equilibrium individually with constant temperatures $T_L$ and $T_R$ respectively, where the side wall of the cylinder is a perfect mirror of zero temperature and no heat is exchanged between the side wall and each lid. Here the side wall is considered just for preparing a radiation field of finite volume. Therefore we ignore any thermodynamic effect of the side wall hereafter. In the cavity case, the inner and outer black bodies are in equilibrium individually with constant temperatures $T_{in}$ and $T_{out}$ respectively. Consequently due to the Stefan-Boltzmann law, a stationary energy flow $J$ exists in both cases. The equilibrium limit is taken with $T_L = T_R$ and $T_{in} = T_{out}$.

The requirement of the zeroth law is not enough to search for consistent definitions of state variables. Therefore we require that the following properties of equilibrium thermodynamics are retained in the SST for a radiation field:
\begin{description}
  \item[(R3)] Any state variable is either intensive or extensive.
  \item[(R4)] The free energy is convex (or concave) with extensive (or intensive) variables.
  \item[(R5)] Thermodynamic functions are related by the Legendre transformation.
  \item[(R6)] The Gibbs-Duhem relation holds to ensure the first law.
  \item[(R7)] Extensive variables are additive.
\end{description}
To make the meaning of requirement (R7) more clear, consider the cylinder case as an example. The internal energy of the left lid, $E_L$, and that of the right lid, $E_R$, are determined by equilibrium thermodynamics, since the lids are in equilibrium individually. Then with the SST internal energy, $E_{rad}$, of a radiation field defined in section \ref{sec-cylinder}, the requirement (R7) denotes that the total energy is given by $E_{tot} = E_L + E_R + E_{rad}$. Similarly, the total entropy is $S_{tot} = S_L + S_R + S_{rad}$, where $S_L$ and $S_R$ are equilibrium entropies of left and right lids, and $S_{rad}$ is the SST entropy defined in section \ref{sec-cylinder}. We require that the same additivity holds for the cavity case. After introducing the idea of local steady states for the cavity case, the additivity of extensive variables of the radiation field in different local steady states is also considered in section \ref{sec-cavity}.

Though the additivity for extensive variables of the radiation field is an appropriate requirement due to the collisionless nature of photons, it should be clarified that the additivity among extensive variables of the two black bodies and the radiation field like $S_{tot} = S_L + S_R + S_{rad}$ is a less appropriate requirement and is nothing but a simple assumption. However, as mentioned at the end of fifth paragraph in section \ref{sec-intro}, we assume the following condition:
\begin{description}
 \item[(C1)] Each black body is always in an equilibrium state, and no fluctuation arises on the radiation field.
\end{description}
Under this restriction, the state variables of the two black bodies are always given by the ordinary thermodynamics (whose extensive variables are additive), and it is appropriate to consider that, once a photon is emitted at a black body, no interaction of the photon with black bodies exists until it is absorbed at a black body. This implies that the interaction between the radiation field and a black body is of a very short range, and that it seems not so bad to assume the additivity like $S_{tot} = S_L + S_R + S_{rad}$. When one proceeds to the study of the fluctuations in our settings which occur in the case when the bodies are in nonequilibrium states, the possibility of existence and influence of non-additive state variables of the bodies should be reconsidered. And when the bodies are so massive that the gravitational effect which is a long range interaction between the bodies can not be ignored, the possibility of existence and influence of non-additive state variables of the bodies should be reconsidered. Further if the gravity becomes so strong that the general relativistic effects can not be ignored, a non-additivity among state variables of the bodies and of the radiation field may arise. And concerning the condition (C1), a practical issue on future improvement is mentioned in section \ref{sec-sd}. 

\begin{figure}[t]
 \begin{center}
 \includegraphics[height=40mm]{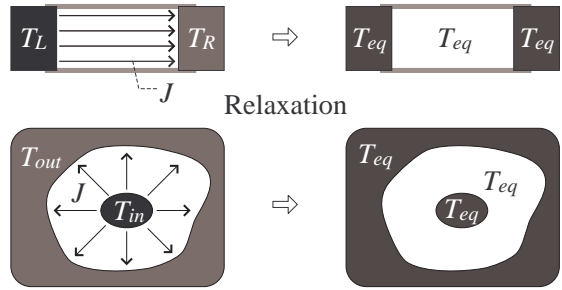}
 \end{center}
\caption{Relaxation process. Total entropy should be non-decreasing, $dS_{tot} \geq 0$.}
\label{pic.2}
\end{figure}

Next in order to extend the second law to steady states, recall the third paragraph in section \ref{sec-intro}. The radiation field in a steady state can never relax to an equilibrium state without introducing interactions with other matters \cite{ref-l.l}. Therefore it is necessary for the second law to consider the relaxation process of the whole system composed of the radiation field and the two black bodies. Here as an example, let us consider again the cylinder case. Cover the cylinder with a heat insulator to be isolated. Then the whole system composed of the two lids and the radiation field relaxes to a total equilibrium state in which the lids and the radiation field have the same equilibrium temperature (figure \ref{pic.2}). If each lid passes a sequence of equilibrium states during this relaxation process, the radiation field passes {\it a sequence of steady states}. The same relaxation process of the whole system in the cavity case is also easily considered. As mentioned at the third paragraph in section \ref{sec-intro}, given a well defined SST entropy $S_{rad}$, the total entropy $S_{tot} \, ( = S_L + S_R + S_{rad} )$ have to be monotone increasing during such a relaxation process of the whole system. Therefore we have to place the other requirement:
\begin{description}
 \item[(R8)] Total entropy is non-decreasing, $dS_{tot} \geq 0$, along the relaxation process of the whole system. And the equality, $dS_{tot} = 0$, holds for a total equilibrium state. 
\end{description}
Though it may appear strange that the non-decreasing nature of the entropy of a nonequilibrium radiation field is not given with a relaxation of an isolated radiation system, such a property should be understood as a unique property to a radiation field. Further, in order to realise the relaxation process of the whole system in a simple way, we require the following condition:
\begin{description}
 \item[(C2)] The time scale for changing the equilibrium states of the two black bodies is sufficiently longer than the flight time of a photon in the space between the two black bodies.
\end{description}
If this condition is not satisfied, the temperatures of the black bodies may change while a photon travels in the space between the two bodies. That is, the effects of retarded time on photons should be taken into account. However the condition (C2) allows us to ignore the retarded time effects.

Further concerning the relaxation process of the whole system, we should pay attention to the special relativistic effect as follows. If the volumes of the two black bodies change during the relaxation process, their surfaces may move. This denotes that the photons are emitted from a moving source and that the special relativistic Doppler effect due to the Lorentz transformation should be considered. However, because the coefficient of thermal expansion of an ordinary solid is small in general, the volumes of the two black bodies do not change so much and the speeds of their surfaces are so small in comparison with the speed of light that the Doppler effect is negligible. Therefore, for simplicity, we assume that the two black bodies satisfy the following condition:
\begin{description}
 \item[(C3)] The two black bodies are {\it rigid} so that their volumes do not change under the influence of the energy flux in the radiation field.
\end{description}

In summary, we aim to construct the steady state thermodynamics for a radiation field in the settings satisfying the conditions (C1), (C2) and (C3) under the requirements from (R1) to (R8).

\section{Cylinder case}\label{sec-cylinder}

\subsection{Energy flux and scale change}
\label{sec-cylinder.energyflux}

The cylinder case is treated in this section. A steady state of a radiation field is accompanied by a stationary energy flow determined by the Stefan-Boltzmann law, $J = \sigma \, \left(\, T_L^{\,4} - T_R^{\,4} \,\right) \, A$, where $A$ is the cross section of the cylinder. Hereafter we set $T_L > T_R$ without loss of generality.

As seen below, the steady state of a radiation field is completely determined by temperatures $T_L$ and $T_R$ without respect to the cross section $A$ and the length of cylinder $L$. This motivates us to restrict the scale change to keep the cylindrical symmetry of the system. That is, the scale change is operated by changing the volume $V \, ( = A L )$ as $V \, \to \, \lambda V$, where $\lambda > 0$ and the scalings of $A$ and $L$ are not uniquely determined. The extensivity and intensivity are defined with this scale change. Then, the volume of the radiation field, $V$, is obviously an extensive variable. However the energy flow $J$ is neither intensive nor extensive, since $J \propto A$ but $J \not\propto L$. Therefore $J$ is not a state variable due to the requirement (R3). On the other hand, when one introduces an energy flux (an energy flow per unit cross-section), $j = \sigma \, \left(\, T_L^{\,4} - T_R^{\,4} \,\right)$, this can be a candidate for a state variable. However it will be shown in subsection \ref{sec-cylinder.tau} that $j$ does not work as a consistent state variable.

\subsection{Distribution function}
\label{sec-cylinder.distribution}

Photons at the moment of emission at a lid have a Planckian distribution of temperature $T_L$ or $T_R$, since the lids of cylinder are in equilibrium due to the condition (C1). Then, because photons are collisionless, the photons emitted at one lid do not alter their Planckian distribution until absorbed at another lid. Hence we can define a distribution function, $d(\vec{p})$, of a steady state for a radiation field as
\sikib
 d(\vec{p}) \equiv \frac{1}{\exp[\hbar \omega / T(\vec{p})] - 1} \, ,
\label{eq-cylinder.distribution}
\sikie
where $\vec{p}$ is a momentum of a photon related to its frequency $\omega = p \, c/\hbar$ and $T(\vec{p})$ is given by
\sikibnon
 T(\vec{p}) =
 \begin{cases}
  T_L & \text{for} \,\,\, \vec{p} = \vec{p}_{L} \\
  T_R & \text{for} \,\,\, \vec{p} = \vec{p}_{R}
 \end{cases} \, ,
\sikienon
where $\vec{p}_L$ is the momentum of a photon emitted at the left lid and $\vec{p}_R$ is that emitted at the right lid (see figure \ref{pic.3}). 

This $d(\vec{p})$ implies that, under the condition (C1), the steady state of a radiation field can be considered as a {\it superposition} of two equilibrium states of temperatures $T_L$ and $T_R$.  Then one may think that there is no meaning in study on the radiation field under the condition (C1). However, note that the superposed radiation field as a whole is in a steady state with the energy flux $j$, and that it has already been revealed that $j$ can not work as a state variable \cite{ref-essex} \cite{ref-wildt}. Therefore, the study of a radiation field under the condition (C1) obviously has two meanings to reveal an appropriate state variable of a nonequilibrium order, and to show an example of a consistent SST, although the new steady state variable and the way of construction of the SST presented in this paper are unique to a radiation field.

\subsection{Internal energy}

\begin{figure}[t]
 \begin{center}
 \includegraphics[height=25mm]{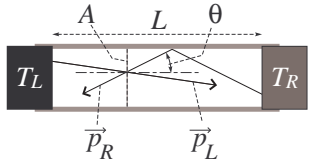}
 \end{center}
\caption{Photons of momentum $\vec{p}_L$ and $\vec{p}_R$. }
\label{pic.3}
\end{figure}

We define an SST internal energy of a radiation field, $E_{rad}$, as the energy carried by all photons,
\sikibnon
 E_{rad} &\equiv&
  2 \int \frac{dp^3}{h^3} \, dx^3 \, \hbar \, \omega \, d(\vec{p})
 =
  \frac{\hbar}{4 \pi^3 c^3}
  \int d\omega d\Omega_p dx^3 \, \omega^3 \, d(\vec{p}) \\
 &=&
  \frac{\hbar}{4 \pi^3 c^3} \, V \,
  \left[ \, \int_{\vec{p}_L} d\omega d\Omega_p \omega^3 \, d(\vec{p}_L)
          + \int_{\vec{p}_R} d\omega d\Omega_p \omega^3 \, d(\vec{p}_R)
  \, \right] \, ,
\sikienon
where the factor 2 in the first line is due to the helicity of a photon, $d\Omega_p$ is the solid-angle element in $\vec{p}$-space, and a relation $\int dp^3/h^3 = (1/8 \pi^3 c^3) \int d\omega d\Omega_p \, \omega^2$ is used. Since the integrands in the last line do not depend on $\Omega_p$, the angular integrals give half of a total solid-angle, $\int_{\vec{p}_L} d\Omega_p = \int_{\vec{p}_R} d\Omega_p = 2 \pi \int_0^{\pi/2} d\theta \, \sin\theta = 2 \pi$, where $\theta$ is an angle shown in figure \ref{pic.3}. Then the SST internal energy $E_{rad}$ becomes
\sikib
 E_{rad} \, = \,
  \frac{2 \, \sigma}{c} \, \left( \, T_L^{\,4} + T_R^{\,4} \, \right) \, V
 \, = \,
  \frac{1}{2} \, \left(\, E_{eq}(T_L,V) + E_{eq}(T_R,V) \,\right) \, ,
\label{eq-cylinder.internal}
\sikie
where $\int_0^{\infty} dx \, x^3 / ( e^x - 1 ) = \pi^4/15$ is used, and $E_{eq}(T,V) = ( 4 \sigma / c ) \, T^4 \, V$ is the internal energy of a radiation field in an equilibrium state of temperature $T$ and volume $V$. This $E_{rad}$ is obviously extensive and satisfies the requirement (R1), $E_{rad} = E_{eq}(T_{eq},V)$ for $T_L = T_R \equiv T_{eq}$. 

\subsection{Entropy}
\label{sec-cylinder.entropy}

To search for a definition of an SST entropy, we try referring \cite{ref-rosen} \cite{ref-ore} and \S55 in \cite{ref-l.l}. In those references, a general form of an entropy of a boson gas in a nonequilibrium state has been derived by making use of the Boltzmann principle. That is, by counting the number of states of bosons and using the Stirling's formula, the entropy of a nonequilibrium boson gas, $S_{boson}$, has been obtained
\sikibnon
 S_{boson} =
  \int \frac{dp^3}{h^3} dx^3 \, g_{\vec{p},\vec{x}}
  \left[\, \left(\, 1 + N_{\vec{p},\vec{x}} \,\right) \,
             \ln\left(\, 1 + N_{\vec{p},\vec{x}} \,\right)
         - N_{\vec{p},\vec{x}} \, \ln N_{\vec{p},\vec{x}}
  \,\right] \, ,
\sikienon
where $g_{\vec{p},\vec{x}}$ and $N_{\vec{p},\vec{x}}$ are respectively the number of states and the average number of bosons at a point $(\vec{p},\vec{x})$ in the phase space of the boson gas. It has also been shown in reference \cite{ref-l.l} that the maximisation of $S_{boson}$ ($\delta S_{boson} = 0$) for an isolated system gives the equilibrium Bose distribution function.

In references \cite{ref-l.l} \cite{ref-rosen} \cite{ref-ore}, concrete forms of $g_{\vec{p},\vec{x}}$ and $N_{\vec{p},\vec{x}}$ are not specified, since an arbitrary system is considered. However for our system, they are determined; $g_{\vec{p},\vec{x}} = 2$ due to the helicity of a photon and $N_{\vec{p},\vec{x}} = d(\vec{p})$. Hence, we define the SST entropy of a radiation field, $S_{rad}$, as
\sikib
 S_{rad} &\equiv&
  2 \int \dfrac{dp^3}{h^3} \, dx^3 \, \left[ \,
      \left(\, 1 + d(\vec{p}) \,\right) \, \ln\left(\, 1 + d(\vec{p}) \,\right) \,
    - d(\vec{p}) \, \ln d(\vec{p})
  \right] \nonumber \\
 &=&
  \frac{8 \, \sigma}{3 \, c} \, \left(\, T_L^{\,3} + T_R^{\,3} \,\right) \, V
 =
  \frac{1}{2} \, \left(\, S_{eq}(T_L,V) + S_{eq}(T_R,V) \,\right) \, ,
\label{eq-cylinder.entropy}
\sikie
where $S_{eq}(T,V) = (16 \sigma / 3 c) \, T^3 \, V$ is the equilibrium entropy, and the following relations are used, $\int_0^{\infty} dx \left[\, x^2/( e^x - 1 ) \, \right]\, \ln( e^x - 1 ) = 11 \pi^4/180$ and $\int_0^{\infty} dx \left[\, x^2/( 1 - e^{-x}) \, \right]\, \ln( 1 - e^{-x} ) = - \pi^4/36$. This $S_{rad}$ is obviously extensive and satisfies the requirement (R1), $S_{rad} = S_{eq}(T_{eq},V)$ for $T_L = T_R \equiv T_{eq}$.

As discussed in section \ref{sec-require}, a well defined entropy has to satisfy the requirement (R8). Indeed as shown in next subsection, $S_{rad}$ of equation (\ref{eq-cylinder.entropy}) satisfies (R8). Therefore $S_{rad}$ of (\ref{eq-cylinder.entropy}) is a candidate for a well defined SST entropy of a radiation field. However, there are many other forms of $S_{rad}$ which can satisfy (R8). Further the reference \cite{ref-k.h-2} conjectures well the inappropriateness of the Shannon-type form like $S_{boson}$ as a nonequilibrium entropy at least for dissipative systems, and hence there is a possibility that $S_{rad}$ of (\ref{eq-cylinder.entropy}) is also not appropriate for the SST entropy of a radiation field (collisionless system). What we can say from above is that, while $S_{rad}$ of (\ref{eq-cylinder.entropy}) satisfies the requirement (R8), it is not clear at present whether or not $S_{rad}$ of (\ref{eq-cylinder.entropy}) is unique and completely appropriate to the SST entropy of a radiation field. Hence, to be exact, we observe whether we can construct a candidate for a consistent SST for a radiation field with $S_{rad}$ of (\ref{eq-cylinder.entropy}).

\subsection{2nd law with a relaxation process}
\label{sec-cylinder.2nd}

\begin{figure}[t]
 \begin{center}
 \includegraphics[height=40mm]{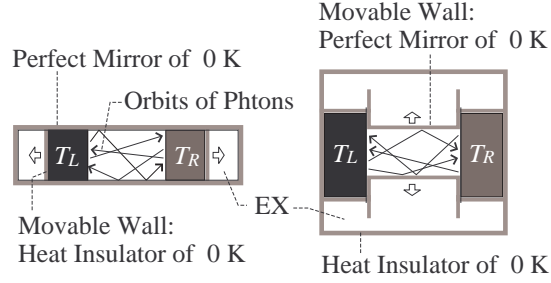}
 \end{center}
\caption{The case of $dV \neq 0$ during the relaxation process. The region denoted as EX is of true vacuum of zero temperature, where even a radiation field does not exist. That is, the smoothly movable wall never give work on the region EX.}
\label{pic.7}
\end{figure}

This subsection shows that $S_{rad}$ of equation (\ref{eq-cylinder.entropy}) satisfies the requirement (R8). We set the cylinder being covered with a heat insulator and consider that each lid passes a sequence of equilibrium states and the radiation field passes a sequence of steady states during the relaxation process of the whole system. Due to the condition (C3), the volumes of the lids does not change during the relaxation. Then the following equilibrium thermodynamic relations hold for the lids,
\sikibnon
 dE_L = C_L \, dT_L = T_L \, dS_L \quad , \quad
 dE_R = C_R \, dT_R = T_R \, dS_R \, ,
\sikienon
where $E_L$, $C_L$ and $S_L$ are the equilibrium internal energy, heat capacity and entropy of the left lid, and $E_R$, $C_R$ and $S_R$ are those of the right lid. Since the whole system is isolated by the heat insulator, the total energy conservation holds,
\sikibnon
 E_{tot}
 &=& E_L + E_R + E_{rad} = \text{const.} \nonumber \\
  \Rightarrow \qquad 0 &=& C_L dT_L + C_R dT_R
   + \frac{8 \, \sigma}{c} \, \left(\, T_L^{\,3} dT_L + T_R^{\,3} dT_R  \,\right) \, V
   + \frac{2 \, \sigma}{c} \, \left(\, T_L^{\,4} + T_R^{\,4} \,\right) \, dV \, .
\sikienon
Here note that, even if the volumes of the lids do not change due to the condition (C3), there is a possibility of the change of the radiation volume $V$ during the relaxation process. Two examples of such a case are shown in figure \ref{pic.7}, where the smoothly movable wall is pushed by the radiation pressure. That is, $dV \geq 0$ during the relaxation. And it is appropriate to assume that the speed of the wall during the relaxation is small enough to ignore the special relativistic Doppler effect on photons and the kinetic energy of the lids. However it should be emphasised that, when the volume $V$ continues to expand, the whole system never reaches any equilibrium state, since the equilibrium state does not evolve in time. Furthermore if the exterior region denoted as EX in figure \ref{pic.7} is sufficiently large, the radiation volume $V$ becomes so large that the condition (C2) will be violated after a sufficiently long time passes. Hence it is appropriate to take the equilibrium limit with $T_L = T_R$ and $dV = 0$ for the relaxation of the whole system. Consequently, the differential of the total entropy $S_{tot} = S_L + S_R + S_{rad}$ becomes
\sikibnon
 dS_{tot}
 &=&
   \frac{C_L}{T_L} \, dT_L + \frac{C_R}{T_R} \, dT_R
 + \frac{8 \, \sigma}{c} \, \left(\, T_L^{\,2} dT_L + T_R^{\,2} dT_R  \,\right) \, V
 + \frac{8 \, \sigma}{3 \,c} \, \left(\, T_L^{\,3} + T_R^{\,3} \,\right) \, dV \\
 &=&
   \left(\, \frac{1}{T_R} - \frac{1}{T_L} \,\right) \,
   \left[ \,
      \left(\, C_R + \frac{8 \, \sigma}{c} \, T_R^{\,3} \, V \,\right) dT_R
    + \frac{2 \, \sigma}{c} \, T_R^{\,4} \, dV \,\right]
 + \frac{2 \, \sigma}{3 \, c} \, \left(\, T_L^{\,3} + T_R^{\,3} \,\right) \, dV \, .
\sikienon
The inequality $dS_{tot} > 0$ holds obviously for $T_L > T_R$ ($dT_R > 0$), and with the help of the energy conservation it is also easily derived that the inequality $dS_{tot} > 0$ holds for the case $T_R > T_L$ ($dT_L > 0$). Further the equality, $dS_{tot} = 0$, holds for the equilibrium limit $T_L = T_R$ and $dV = 0$. As mentioned at the requirement (R8) in section \ref{sec-require}, this denotes that $S_{rad}$ of equation (\ref{eq-cylinder.entropy}) is a candidate for a well defined SST entropy. 

\subsection{Pressure}
\label{sec-cylinder.pressure}

We define an SST pressure by the tensor,
\sikib
 \pt_{\mu \nu} \equiv
    2 \int \frac{dp^3}{h^3} \, \frac{p}{c} \, c_{\mu} \, c_{\nu} \, d(\vec{p}) \, ,
\label{eq-cylinder.pressure}
\sikie
where the factor $2$ is due to the helicity of a photon, $c_{\mu}$ is a spatial component of a photon velocity satisfying $\Sigma_{\mu} c_{\mu}^{\,2} = c^2$, and $\mu$ and $\nu$ denote the Cartesian components. This tensor $\pt$ corresponds to the pressure tensor of an ordinary dissipative system, since $p/c$ corresponds to the mass of a molecule. $\pt$ is obviously symmetric, $\pt_{\mu \nu} = \pt_{\nu \mu}$. 

At the equilibrium limit $T_L = T_R \equiv T_{eq}$, the SST pressure tensor becomes
\sikibnon
 \pt =
  \frac{1}{3} \, \text{tr}\pt \, \delta_{\mu \nu}  \quad , \quad
 \text{tr}\pt =
  2 \int \frac{dp^3}{h^3} \, \frac{p c}{\exp(\hbar \omega/T_{eq}) - 1}
 = \frac{4\, \sigma}{c} \, T_{eq}^{\,4} = 3 \, P_{eq}(T_{eq}) \, ,
\sikienon
where $p c = \hbar \omega$ is used, tr$\pt$ is the trace of $\pt$ and $P_{eq}(T) = ( 4 \sigma/3 c ) T^4$ is the equilibrium pressure. This denotes the SST pressure tensor $\pt$ satisfies the requirement (R1).

\subsection{Free energy}
\label{sec-cylinder.free}

We try to define an SST free energy, $F_{rad}$, as a scalar variable satisfying the following equilibrium limit,
\sikibnon
 \left. - \pd{F_{rad}}{V} \right|_{T_L = T_R = T_{eq}} = P_{eq}(T_{eq}) \, .
\sikienon
Therefore, since $V$ is also a scalar variable, we are to require
\sikib
 - \pd{F_{rad}}{V} = P_{rad} \, ,
\label{eq-cylinder.condition.free}
\sikie
where $P_{rad}$ is the scalar quantity which is constructed using only the SST pressure tensor $\pt$ and reduces to $P_{eq}$ at equilibrium limits. This scalar $P_{rad}$ can be given by the trace of $\pt$ as
\sikib
 P_{rad} 
 = \frac{1}{3}\, \text{tr}\pt
 = \frac{2}{3} \int\frac{dp^3}{h^3} \, \hbar \omega \, d(\vec{p})
 = \frac{1}{3} \, \frac{E_{rad}}{V}
 = \frac{2\, \sigma}{3\, c} \,
   \left( T_L^{\,4} + T_R^{\,4} \right) \, .
\label{eq-cylinder.trace}
\sikie
It is obvious that $P_{rad} = P_{eq}(T_{eq})$ for $T_L = T_R \equiv T_{eq}$. As will be seen later, this definition of the scalar $P_{rad}$ ensures the consistency of our SST formulation. Hence we define $F_{rad}$ as
\sikib
 F_{rad} \equiv
   - \frac{1}{3}\, \text{tr}\pt \, V
 = - \frac{2 \, \sigma}{3 \, c} \, \left(\, T_L^{\,4} + T_R^{\,4} \,\right) \, V
 = \frac{1}{2} \, \left(\, F_{eq}(T_L,V) + F_{eq}(T_R,V) \,\right) \, ,
\label{eq-cylinder.free}
\sikie
where $F_{eq}(T,V) = - (4 \sigma / 3 c) \, T^4 \, V$ is the equilibrium free energy. This $F_{rad}$ is obviously extensive and satisfies the requirement (R1), $F_{rad} = F_{eq}(T_{eq},V)$ for $T_L = T_R \equiv T_{eq}$. A related discussion is given in section \ref{sec-sd}.

Note that, with only the relation (\ref{eq-cylinder.condition.free}), an arbitrary function of $T_L$ and $T_R$ can be added to the definition of $F_{rad}$ as $F_{rad} \equiv - (1/3)\,$tr$\pt \, V + K(T_L,T_R)$. However $F_{rad}$ should be an extensive variable due to the requirement (R3). Consequently, the arbitrary function disappears, $K = 0$.

\subsection{Chemical potential}
\label{sec-cylinder.chemical}

The chemical potential in general can be interpreted as a work needed to add a particle to a system under consideration. Because a photon is a collisionless particle, no work is required to add a new photon into the radiation field in the cylinder. This is the case either for equilibrium states or for steady states. Indeed, the chemical potential of a radiation field in equilibrium is zero. Therefore it is appropriate to set the SST chemical potential of a radiation field also as zero.

\subsection{Intensive nonequilibrium order parameter}
\label{sec-cylinder.tau}

As a candidate for a nonequilibrium order parameter of a steady state, we can recommend the energy flow $J$, its areal density $j = \sigma \, ( T_L^{\,4} - T_R^{\,4})$ and the temperature difference,
\sikib
 \tau \equiv T_L - T_R \, ,
\label{eq-cylinder.temperature.diff}
\sikie
where $\tau$ is defined as a positive quantity due to the assumption $T_L > T_R$ .

Note that, as mentioned at subsection \ref{sec-cylinder.energyflux}, $J$ is not a state variable. Therefore $J$ is excluded from a state variable describing a nonequilibrium order. Next look at the energy flux $j$. If $j$ is a consistent state variable, its conjugate variable should also be defined in a consistent way. However the following calculation shows that the conjugate variable to $j$ disappears for any value of $j$,
\sikib
 \pd{F_{rad}}{j}
 = - \frac{8 \, \sigma}{3 \, c} \,
     \left(\, T_L^{\,3}\, \pd{T_L}{j} + T_R^{\,3}\, \pd{T_R}{j} \,\right) \, V
 = - \frac{8 \, \sigma}{3 \, c} \,
     \left(\, T_L^{\,3}\, \frac{1}{4 \, \sigma \, T_L^{\,3}}
            - T_R^{\,3}\, \frac{1}{4 \, \sigma \, T_R^{\,3}} \,\right) \, V = 0 \, .
\label{eq-cylinder.j.conjugate}
\sikie
This means that the conjugate variable to $j$ can not be defined in a consistent way, and that $j$ should also be excluded from a state variable describing a nonequilibrium order. 

On the other hand, $\tau$ is obviously intensive and satisfies the requirement (R2), $\tau = 0$ for $T_L = T_R$. Thus we adopt $\tau$ as the intensive state variable for a nonequilibrium order of steady states.

\subsection{Temperature and extensive nonequilibrium order parameter}
\label{sec-cylinder.temperature.psi}

We proceed to search for a definition of an SST temperature, $T_{rad}$. One may naively try to use the relation, $\partial S_{rad}/\partial E_{rad} = 1/T_{rad}$, to obtain the form of $T_{rad}$. In ordinary equilibrium thermodynamics for a radiation field, this partial derivative is calculated by fixing the extensive variable $V$. For our system, an extensive state variable of a nonequilibrium order of steady states should also be one of the fixed variables in the partial derivative. However, because such an extensive variable has not been specified yet, we can not evaluate the partial derivative here. Therefore we try to search for the form of $T_{rad}$ with the other relation,
\sikibnon
 - \left( \pd{F_{rad}}{T_{rad}} \right)_{V , \tau} = S_{rad} \, .
\sikienon
Further we require that the temperatures $T_L$ and $T_R$ depend on $T_{rad}$ and $\tau$,
\sikibnon
 T_L = T_L(T_{rad},\tau) \quad , \quad T_R = T_R(T_{rad},\tau) \, .
\sikienon
Then using the form of $S_{rad}$ and $F_{rad}$ we find, 
\sikibnon
 - \pd{F_{rad}}{T_{rad}} = S_{rad} \quad \Rightarrow \quad
 \pd{T_L}{T_{rad}} = 1 \quad , \quad \pd{T_R}{T_{rad}} = 1 \, .
\sikienon
Therefore, with the relation $\tau = T_L - T_R$, we can set
\sikibnon
 T_L = T_{rad} + \alpha \, \tau + L(\tau) \quad , \quad
 T_R = T_{rad} - ( 1 - \alpha ) \, \tau + L(\tau) \, ,
\sikienon
where $\alpha$ is a constant and $L(\tau)$ is a function of $\tau$ determined below. These temperatures give the form of $T_{rad}$ as
\sikib
 T_{rad} = ( 1 - \alpha ) \, T_L + \alpha \, T_R - L(\tau) \, .
\label{eq-cylinder.alpha.L}
\sikie

For later use to determine the form of $L(\tau)$, we define an extensive nonequilibrium order parameter, $\Psi$, as
\sikibnon
 \Psi
 \equiv - \pd{F_{rad}}{\tau}
 = \frac{8 \, \sigma}{3 \, c} \,
   \left(\, \left(\, \alpha + L' \,\right) \, T_L^{\,3}
          + \left(\, - ( 1 - \alpha ) + L' \,\right) \, T_R^{\,3} \,\right) \, V \, ,
\sikienon
where $L'(\tau) = dL(\tau)/d\tau$. The requirement (R2) gives the relation,
\sikibnon
 \Psi(T_L = T_R) = 0 \quad \Rightarrow \quad L'(0) = \frac{1}{2} - \alpha \, .
\sikienon
Further the Legendre transformation should hold due to the requirement (R5),
\sikib
 E_{rad} = F_{rad}(T_{rad} , \tau , V) + T_{rad} \, S_{rad} + \tau \, \Psi \, ,
\label{eq-cylinder.legendre}
\sikie
where $F_{rad}$ is treated as a function of $T_{rad}$, $\tau$ and $V$. Here note that, since $\tau$ has the dimension of temperature and $\Psi$ has that of entropy, the signature of the term $\tau \, \Psi$ in equation (\ref{eq-cylinder.legendre}) is positive. Substituting $\Psi$ and $T_{rad}$ of (\ref{eq-cylinder.alpha.L}) into (\ref{eq-cylinder.legendre}), we obtain
\sikibnon
 L' \, T_L - L' \, T_R = L \quad \Rightarrow \quad
 L'(\tau) = \frac{L(\tau)}{\tau} \quad \Rightarrow \quad
 L(\tau) = \lambda \, \tau \, ,
\sikienon
where $\lambda$ is a constant. 
Consequently we find $\alpha + \lambda = 1/2$, and $T_{rad}$ of the form (\ref{eq-cylinder.alpha.L}) becomes
\sikib
 T_{rad} = \frac{1}{2} \, \left(\, T_L + T_R \,\right) \, .
\label{eq-cylinder.temperature}
\sikie
The extensive nonequilibrium order parameter becomes
\sikib
 \Psi =
   \frac{4 \, \sigma}{3\, c} \, \left(\, T_L^{\,3} - T_R^{\,3} \,\right) \, V
 = \frac{1}{4} \, \left(\, S_{eq}(T_L,V) - S_{eq}(T_R,V) \, \right) \, .
 \label{eq-cylinder.entropy.diff}
\sikie
The form of $T_{rad}$ denotes that the third law holds in the SST for a radiation field if the third law of the equilibrium thermodynamics is retained for the left and right lids.

\subsection{Consistency of thermodynamic functions and 1st law}

From above results, it can be checked easily that the requirement (R4) is satisfied,
\sikibnon
 &&
 \pdd{F_{rad}}{V} = 0 \, (\geq 0) \,\, \text{: convex} \quad / \quad
 \pdd{F_{rad}}{T_{rad}} \leq 0 \quad , \quad
 \pdd{F_{rad}}{\tau} \leq 0 \,\, \text{: concave} \, .
\sikienon
Further, we find that the requirement (R6) is satisfied,
\sikib
 &&
 - S_{rad} \, dT_{rad} + V\, dP_{rad} - \Psi \, d\tau = 0 \, ,
 \label{eq-cylinder.gibs.duhem.diff} \\
 &&
 E_{rad} = T_{rad} \, S_{rad} - P_{rad} \, V + \tau \, \Psi \, ,
 \label{eq-cylinder.gibs.duhem}
\sikie
where $P_{rad}$ is given by the SST pressure tensor $\pt$ as equation (\ref{eq-cylinder.trace}). These are the Gibbs-Duhem relations. Then we obtain the first law,
\sikib
 dE_{rad} = T_{rad} \, dS_{rad} - P_{rad} \, dV + \tau \, d\Psi \, .
\label{eq-cylinder.1st}
\sikie

Here note that, for example for an ordinary continuum system, the work term in the first law is given like $P_{i j} \, d\epsilon_{i j}$, where $P_{i j}$ is the stress tensor (which corresponds to the pressure tensor) and $\epsilon_{i j}$ is the strain tensor. The tensor $\epsilon_{i j}$ can be decomposed, $\epsilon_{i j} = (1/3)\,$tr$(\epsilon) \, \delta_{i j} + \epsilon_{i j}^{(c)}$, where the trace tr$(\epsilon)$ is the bulk strain and $\epsilon_{i j}^{(c)}$ is the constant volume strain, tr$(\epsilon^{(c)}) = 0$. That is, for an ordinary continuum system, the work term becomes
\sikibnon
 P_{i j}\, d\epsilon_{i j} = \frac{1}{3} \text{tr}P \, dV + P_{i j} \, d\epsilon_{i j}^{(c)} \, ,
\sikienon
where $dV = d[$tr$(\epsilon) ]$ means the change of volume. With this observation, we can discuss the first law (\ref{eq-cylinder.1st}) as follows. In the context of the cylinder case, the constant volume strain means the change of the cross section $A$ and the length $L$ of the cylinder as $A \to \lambda \, A$ and $L \to L/\lambda$, which denotes $V$ does not change. Then, since all of the state variables obtained in this section \ref{sec-cylinder} are characterised by $T_L$, $T_R$ and $V$, the constant volume strain can not cause a change of state variables like $E_{rad}$. That is, no interaction among photons in a radiation field causes zero energy change in the radiation field by any deformation of the system so long as the whole volume of the radiation field $V$ does not change.  Hence it is reasonable that the constant volume strain does not contribute to the first law (\ref{eq-cylinder.1st}) and only the bulk strain appears in the work term as $- P_{rad} \, dV$ in (\ref{eq-cylinder.1st}). The absence of the work term of the constant volume strain should be understood as a unique property of a radiation field. More discussions are given in section \ref{sec-sd}. 

Finally, it follows that
\sikibnon
 \left( \pd{S_{rad}}{E_{rad}} \right)_{V,\Psi} = \frac{1}{T_{rad}} \, .
\sikienon
This denotes that our definition of $T_{rad}$, $S_{rad}$, $E_{rad}$ and $\Psi$ are theoretically reasonable. 

As seen so far, the thermodynamic formulation (0th, 1st, 2nd and 3rd laws) can be obtained for the steady states of a radiation field in the cylinder case.

\section{Cavity case}\label{sec-cavity}

\subsection{Local steady state}

The cavity case (right one in figure \ref{pic.1}) is treated in this section. The geometrical shapes of the inner black body and the cavity are arbitrarily set. For example, consider a case that the inner black body is a thin board, and count the number of photons coming from the board to a spatial point. Such a photon number at a point in front of the surface of the board is larger than that at a point on the plane of the board. Thus, in general, it is recognised that the number of photons coming from the inner black body to a point varies from point to point. The same is true of the number of photons coming from the outer black body. This means that the distribution function of photons is anisotropic and inhomogeneous as shown in next subsection. On the other hand, for example, the anisotropy and inhomogeneity of the velocity field of an ordinary fluid system have been traditionally handled well with the idea of local equilibrium in the ordinary fluid dynamics. Hence, after such a success of the idea of {\it locality} for the anisotropic and inhomogeneous system, we adopt the idea of {\it local steady states}. The radiation field in a sufficiently small region is in a steady state, but the steady state in one small region may be different from that in the other small region. Therefore the state variable in the cavity case should be defined as a function of a spatial point $\vec{x}$, where the extensive variable is to be understood as a density. 

The idea of local steady states is not unique for a radiation field, but it also appears in formulating the steady state thermodynamics for dissipative systems \cite{ref-s.t}. It is the inhomogeneity of the temperature of the matter that raises the idea of local steady states in reference \cite{ref-s.t}. On the other hand, it has already been revealed by reference \cite{ref-k.h-1} that, for a steady state of a hard core molecule gas system, the inhomogeneity of the temperature causes the inhomogeneity of the distribution function. Therefore the origin of the idea of local steady states in reference \cite{ref-s.t} seems to be the same as that in our radiation system. Further in the extended irreversible thermodynamics \cite{ref-eit}, the idea of {\it local nonequilibrium state} is suggested, since, for example, some state variables are treated as functions of spatial point. Therefore, the idea of locality seems to be universal without respect to the nature of interactions among components of the nonequilibrium system under consideration, as though the formalism presented in this paper, for example the exclusion of the energy flux from state variables, is unique to steady states of a radiation field. 

\begin{figure}[t]
 \begin{center}
 \includegraphics[height=20mm]{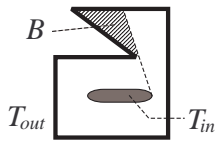}
 \end{center}
\caption{The region $B$ is excluded from the volume of a radiation field in local steady states.}
\label{pic.4}
\end{figure}

Before proceeding to the search for definitions of state variables, let us note about the spatial volume filled with a radiation field. Because photons are collisionless, the region $B$ shown in figure \ref{pic.4} should be excluded from the volume, $V_{rad}$, of a radiation field which is in local steady states. The radiation field in the region $B$ is in an equilibrium state of temperature $T_{out}$. 

\subsection{Distribution function}

Let $\vec{p}_{in}$ and $\vec{p}_{out}$ denote the momentums of photons emitted at the inner black body and at the outer black body respectively. The directions from which the photons of $\vec{p}_{in}$ and $\vec{p}_{out}$ can come to a spatial point $\vec{x}$ vary from point to point. Then the distribution function in the cylinder case (\ref{eq-cylinder.distribution}) is extended to the cavity case as
\sikib
 d(\vec{x},\vec{p}) \equiv \frac{1}{\exp[\hbar \omega / T(\vec{x},\vec{p})] - 1} \, ,
\label{eq-cavity.distribution}
\sikie
where $\omega = p c / \hbar$, and $T(\vec{x},\vec{p})$ is given by
\sikibnon
 T(\vec{x},\vec{p}) =
 \begin{cases}
  T_{in}  & \text{for} \,\,\, \vec{p} = \vec{p}_{in} \,\,\, \text{at $\vec{x}$} \\
  T_{out} & \text{for} \,\,\, \vec{p} = \vec{p}_{out} \,\,\, \text{at $\vec{x}$}
 \end{cases} \, .
\sikienon

\subsection{Internal energy density}

\begin{figure}[t]
 \begin{center}
 \includegraphics[height=25mm]{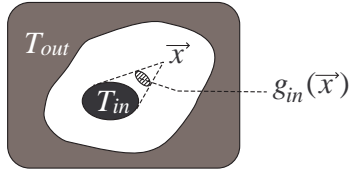}
 \end{center}
\caption{The geometrical factor, $\gin{x}$, is shown. Another one is $\gout{x} = 1 - \gin{x}$.}
\label{pic.5}
\end{figure}

The SST internal energy density, $\erad{x}$, is defined as the energy of photons at a point $\vec{x}$,
\sikibnon
 \erad{x} \equiv
   2 \int \frac{dp^3}{h^3} \, \hbar \, \omega \, d(\vec{x},\vec{p})
 = \frac{\sigma}{\pi \, c} \,
   \left[ \, \int_{\vec{p}_{in} } d\Omega_{xp} \, T_{in }^{\,4}
           + \int_{\vec{p}_{out}} d\Omega_{xp} \, T_{out}^{\,4}
   \, \right] \, ,
\sikienon
where the factor 2 in the first line is due to the helicity of a photon, $d\Omega_{xp}$ is the solid-angle element in $\vec{p}$-space at a spatial point $\vec{x}$. Because the solid-angle of the inner black body measured from a spatial point $\vec{x}$ takes on different values according to the position of $\vec{x}$, the numerical values of the integrals in the last line vary from point to point. Therefore we introduce the geometrical factors, $g_{in}(\vec{x})$ and $g_{out}(\vec{x})$, as
\sikib
 \gin{x}  \equiv \frac{1}{4 \pi} \, \int_{\vec{p}_{in} } d\Omega_{xp} \quad , \quad
 \gout{x} \equiv \frac{1}{4 \pi} \, \int_{\vec{p}_{out}} d\Omega_{xp} \, .
\label{eq-cavity.geometrical}
\sikie
One of them, $\gin{x}$, is shown in figure \ref{pic.5}. These geometrical factors satisfy a relation, $\gin{x} + \gout{x} = 1$, and are positive, $\gin{x} > 0$ and $\gout{x} > 0$. Then the SST internal energy density becomes
\sikib
 \erad{x}
 = \frac{4 \, \sigma}{c} \,
   \left(\, \gin{x} \, T_{in}^{\,4} + \gout{x} \, T_{out}^{\,4} \,\right)
 = \gin{x} \, e_{eq}(T_{in}) + \gout{x} \, e_{eq}(T_{out}) \, ,
\label{eq-cavity.internal}
\sikie
where $e_{eq}(T) = (4 \sigma / c) \, T^4$ is the equilibrium internal energy density of temperature $T$. This $\erad{x}$ satisfies the requirement (R1), $e_{rad} = e_{eq}(T_{eq})$ for $T_{in} = T_{out} \equiv T_{eq}$. Further due to the requirement (R7), a total internal energy of a radiation field, $E_{rad}$, is given by a volume integral of $\erad{x}$,
\sikibnon
 E_{rad} = \int_{V_{rad}} dx^3 \, \erad{x} \, ,
\sikienon
where the region like $B$ in figure \ref{pic.4} should be excluded from the integral region. It is obvious that the cylinder case is recovered by setting $\gin{x} = \gout{x} = 1/2$.

\subsection{Entropy density}

As for the cylinder case, we try referring \cite{ref-rosen} \cite{ref-ore} and \S55 in \cite{ref-l.l} to define the SST entropy density, $\srad{x}$, as
\sikib
 \srad{x} &\equiv&
   2 \int \dfrac{dp^3}{h^3} \, \left[ \,
       \left(\, 1 + d(\vec{x},\vec{p}) \,\right) \, \ln\left(\, 1 + d(\vec{x},\vec{p}) \,\right) \,
     - d(\vec{x},\vec{p}) \, \ln d(\vec{x},\vec{p})
   \right] \nonumber \\
 &=&
   \frac{16 \, \sigma}{3 \, c} \,
   \left(\, \gin{x}\, T_{in}^{\,3} + \gout{x}\, T_{out}^{\,3} \,\right)
 = \gin{x}\, s_{eq}(T_{in}) + \gout{x}\, s_{eq}(T_{out}) \, ,
\label{eq-cavity.entropy}
\sikie
where $s_{eq}(T) = (16 \sigma / 3 c) \, T^3$ is the equilibrium entropy density. This $\srad{x}$ satisfies the requirement (R1), $\srad{x} = s_{eq}(T_{eq})$ for $T_{in} = T_{out} \equiv T_{eq}$. Due to the requirement (R7), a total entropy of a radiation field, $S_{rad}$, is given by a volume integral of $\srad{x}$,
\sikibnon
 S_{rad} = \int_{V_{rad}} dx^3 \, \srad{x} \, .
\sikienon
The cylinder case is recovered by setting $\gin{x} = \gout{x} = 1/2$. As discussed at the last paragraph in subsection \ref{sec-cylinder.entropy}, $\srad{x}$ is just one candidate for a well defined SST entropy density. Hence, to be exact, we observe whether we can construct a candidate for a consistent SST for a radiation field with $\srad{x}$ of equation (\ref{eq-cavity.entropy}).

\subsection{2nd law with a relaxation process}
\label{sec-cavity.2nd}

This subsection shows that $\srad{x}$ of equation (\ref{eq-cavity.entropy}) satisfies the requirement (R8). We set the outer black body being covered with a heat insulator and consider that each black body passes a sequence of equilibrium states and the radiation field passes a sequence of steady states during the relaxation process of the whole system. Due to the conditions (C1) and (C3), the ordinary thermodynamic relations hold for the black bodies, and the volumes of the two black bodies does not change. Further due to the requirement (R7), the total energy $E_{rad}$ and the total entropy $S_{rad}$ are given by $E_{tot} = E_{in} + E_{out} + E_{rad}$ and $S_{tot} = S_{in} + S_{out} + S_{rad}$ respectively, where $E_{in}$ and $S_{in}$ are the equilibrium internal energy and entropy of the inner black body, and $E_{out}$ and $S_{out}$ are those of the outer black body. 

Here it should be noted that the inner black body may move due to the radiation pressure. However, because of the condition (C3), the radiation volume $V_{rad}$ does not change, $dV_{rad} = 0$. Further it is appropriate to assume that the velocity of the inner black body is very small in comparison with the speed of light. Hence the special relativistic Doppler effect and the kinetic energy of the inner black body can be ignored in the analysis of the relaxation process. And it should also be pointed out that the inner black body rests somewhere in the cavity when the whole system is in a total equilibrium state.

With the aid of the energy conservation and $dV_{rad} = 0$, a calculation similar to the one done in subsection \ref{sec-cylinder.2nd} gives the following inequality,
\sikibnon
 dS_{tot}
 &=&
   \frac{C_{in}}{T_{in}} \, dT_{in} + \frac{C_{out}}{T_{out}} \, dT_{out}
 + \frac{16 \, \sigma}{c} \,
    \left(\, G_{in} \, T_{in}^{\,2} dT_{in}
           + G_{out}\, T_{out}^{\,2} dT_{out}  \,\right) \\
 &=&
   \left(\, \frac{1}{T_{out}} - \frac{1}{T_{in}} \,\right) \,
   \left(\, C_{out} + \frac{16 \, \sigma}{c} \, G_{out} \, T_{out}^{\,3} \right) dT_{out} \geq 0 \, ,
\sikienon
where $G_{in} = \int_{V_{rad}} dx^3 \, \gin{x}$, $G_{out} = \int_{V_{rad}} dx^3 \, \gout{x}$, and $C_{in}$ and $C_{out}$ are the heat capacities of the inner and outer black bodies respectively. The inequality, $dS_{tot} > 0$, holds for both cases $T_{in} > T_{out}$ and $T_{in} < T_{out}$, and the equality, $dS_{tot} = 0$, holds for the equilibrium case $T_{in} = T_{out}$. This indicates that $\srad{x}$ of equation (\ref{eq-cavity.entropy}) is a candidate for a well defined SST entropy.

\subsection{Pressure}

One may naively expect that the pressure of a {\it steady} state is a global quantity, since the pressure gradient in an ordinary dissipative system accelerates the components of the system to cause a dynamical evolution in the system. However, since no dissipation exists among photons, this discussion can not be applied to a radiation field. As seen below, the spatial dependence of $d(\vec{x},\vec{p})$ makes the SST pressure tensor for the cavity case a function of $\vec{x}$.

As for the cylinder case, we define the SST pressure by the tensor,
\sikib
 \pt_{\mu \nu}(\vec{x})
 = 2 \int \frac{dp^3}{h^3} \, \frac{p}{c} \, c_{\mu} \, c_{\nu} \, d(\vec{x},\vec{p}) \, .
\label{eq-cavity.pressure}
\sikie
It can be checked by the same calculation in subsection \ref{sec-cylinder.pressure} that this $\pt(\vec{x})$ satisfies the requirement (R1).

\subsection{Free energy density}

The discussion of the free energy given in subsection \ref{sec-cylinder.free} is also applied for the cavity case. Therefore, since we are searching for a density of an extensive variable, the SST free energy density, $\frad{x}$, is defined as
\sikibnon
 \frad{x} \equiv - \frac{1}{3} \, \text{tr}\pt(\vec{x})
 = - \frac{2}{3}\, \int \frac{dp^3}{h^3} \, \hbar \omega \, d(\vec{x},\vec{p})
 = - \frac{1}{3} \, \erad{x} \, .
\sikienon
Hence we obtain
\sikib
 \frad{x}
 = - \frac{4 \, \sigma}{3 \, c} \,
     \left(\, \gin{x} \, T_{in}^{\,4}
            + \gout{x}\, T_{out}^{\,4} \,\right)
 = \gin{x} \, f_{eq}(T_{in}) + \gout{x} \, f_{eq}(T_{out}) \, ,
\label{eq-cavity.free}
\sikie
where $f_{eq}(T) = - (4 \sigma / 3 c) \, T^4$ is the equilibrium free energy density. This $\frad{x}$ satisfies the requirement (R1), $f_{rad} = f_{eq}(T_{eq})$ for $T_{in} = T_{out} \equiv T_{eq}$. The cylinder case is recovered by setting $\gin{x} = \gout{x} = 1/2$ and integrating $\frad{x}$ over $V_{rad}$.

\subsection{Chemical potential}

The discussion of the chemical potential given at subsection \ref{sec-cylinder.chemical} is also true of the cavity case. Hence we set the SST chemical potential to be zero in the cavity case, too.

\subsection{Intensive nonequilibrium order parameter}

There are three candidates for a nonequilibrium order parameter; the temperature difference $\tau \equiv T_{in} - T_{out}$, the energy flow $J = \sigma \, ( T_{in}^{\,4} - T_{out}^{\,4} ) \, A_{in}$, where $A_{in}$ is the surface area of the inner black body, and the third candidate is the areal density of $J$ at the surface of the inner black body, $j_{in} = \sigma \, ( T_{in}^{\,4} - T_{out}^{\,4} )$. If $J$ and $j_{in}$ are adopted as state variables, it is impossible to recover the cylinder case by setting $\gin{x} = \gout{x} = 1/2$, since $J$ and $j_{in}$ in the cylinder case are not state variables. Therefore it is also appropriate for the cavity case to adopt $\tau$ as a state variable of a nonequilibrium order. Here $\tau$ is intensive in the cavity case, too. More discussion on the definition of $\tau$ is given in section \ref{sec-sd}.

\subsection{Temperature and extensive nonequilibrium order parameter}

To find a definition of the SST temperature, $\Trad{x}$, we follow the same discussion and procedure given in subsection \ref{sec-cylinder.temperature.psi}. The same calculations until (\ref{eq-cylinder.alpha.L}) hold for the cavity case except for that the constant $\alpha$ and the function $L(\tau)$ are modified to depend on a spatial point, $\alpha(\vec{x})$ and $L(\vec{x} ; \tau)$. Then we define an extensive state variable, $\psirad{x}$, which is conjugate to $\tau$,
\sikibnon
 \psirad{x} \equiv
  - \pd{\frad{x}}{\tau}
 = \frac{16 \, \sigma}{3 \, c} \, 
   \left(\, \left(\, \alpha + L' \,\right) \, g_{in} \, T_{in}^{\,3}
          + \left(\, - ( 1 - \alpha ) + L' \,\right) \, g_{out} \, T_{out}^{\,3} \,\right) \, ,
\sikienon
where $L' = \partial L(\vec{x};\tau)/\partial \tau$. The requirement (R2) gives the relation,
\sikibnon
 \psirad{x} = 0 \,\,\, \text{for} \,\,\, T_{in} = T_{out} \quad \Rightarrow \quad
 L'(\vec{x};0) = \gout{x} - \alpha(\vec{x}) \, ,
\sikienon
where $g_{in} + g_{out} = 1$ is used. Further due to the requirement (R5), the Legendre transformation (\ref{eq-cylinder.legendre}) should also hold in the cavity case, 
\sikib
 \erad{x} = f_{rad}(\vec{x} ; \Trad{x},\tau) + \Trad{x} \, \srad{x} + \tau \, \psirad{x} \, .
\label{eq-cavity.legendre}
\sikie
Substituting $\psirad{x}$ and $T_{rad}$ of the form (\ref{eq-cylinder.alpha.L}) into (\ref{eq-cavity.legendre}), we obtain
\sikibnon
 L' \, T_{in} - L' \, T_{out} = L \quad \Rightarrow \quad
 L(\vec{x};\tau) = \lambda(\vec{x}) \, \tau \, ,
\sikienon
where $\lambda$ is an unknown function. Consequently we find $\alpha(\vec{x}) + \lambda(\vec{x}) = \gout{x}$, and $T_{rad}$ of form (\ref{eq-cylinder.alpha.L}) becomes
\sikib
 \Trad{x} = \gin{x} \, T_{in} + \gout{x} \, T_{out} \, .
\label{eq-cavity.temperature}
\sikie
The nonequilibrium order parameter $\psirad{x}$ becomes
\sikib
 \psirad{x}
 = \frac{16 \, \sigma}{3 \, c} \, \gin{x} \, \gout{x} \,
   \left(\, T_{in}^{\,3} - T_{out}^{\,3} \,\right)
 = \gin{x} \, \gout{x} \, \left(\, s_{eq}(T_{in}) - s_{eq}(T_{out}) \, \right) \, .
 \label{eq-cavity.entropy.diff}
\sikie
The cylinder case is recovered by setting $\gin{x} = \gout{x} = 1/2$ and integrating $\psirad{x}$ over $V_{rad}$. The form of $\Trad{x}$ shows that the third law holds in the SST for a radiation field if the third law of the equilibrium thermodynamics is retained for the inner and outer black bodies.

\subsection{Consistency of thermodynamic functions and 1st law}

From the above results, the requirement (R4) is automatically satisfied,
\sikibnon
 \pdd{\frad{x}}{\Trad{x}} \leq 0 \quad , \quad
 \pdd{\frad{x}}{\tau} \leq 0 \,\, \text{: concave} \, .
\sikienon
Further the requirement (R6) is satisfied,
\sikib
 &&
 \left.
  - \srad{x} \, d\Trad{x}
  + d\Prad{x}
  - \psirad{x} \, d\tau \,\,
 \right|_{\vec{x}=\text{fixed}}
 \,=\, 0 \, ,  \label{eq-cavity.gibs.duhem.diff} \\
 &&
 \erad{x} = \Trad{x} \, \srad{x} - \Prad{x} + \tau \, \psirad{x} \, ,
 \label{eq-cavity.gibs.duhem}
\sikie
where
\sikibnon
 \Prad{x}
 = \frac{1}{3} \, \text{tr}\pt(\vec{x})
 = \frac{4\, \sigma}{3\, c} \,
   \left(\, \gin{x}\, T_{in}^{\,4} + \gout{x}\, T_{out}^{\,4} \,\right) \, .
\sikienon
Then the first law is obtained,
\sikib
 \left. d\erad{x} \,\, \right|_{\vec{x}=\text{fixed}}
 \,=\,
 \left.
  \Trad{x} \, d\srad{x} + \tau \, d\psirad{x} \,\, \right|_{\vec{x}=\text{fixed}} \, .
\label{eq-cavity.first}
\sikie
Finally it follows
\sikibnon
 \left(\pd{\srad{x}}{\erad{x}}\right)_{\psirad{x} , \vec{x}} = \frac{1}{\Trad{x}} \, .
\sikienon
This indicates that $\Trad{x}$, $\srad{x}$, $\erad{x}$ and $\psirad{x}$ is theoretically reasonable.

As seen so far, the thermodynamic formulation (0th, 1st, 2nd and 3rd laws) can be extended to the steady states of a radiation field in the cavity case.

\section{Summary and discussions}\label{sec-sd}

{\bf Summary}: 
Based on the collisionless nature of photons, we have been able to find the distribution function of photons in a two-temperature steady state. Consequently, using the Shannon-type entropy (\ref{eq-cylinder.entropy}), a candidate for a consistent two-temperature steady state thermodynamics (SST) for a radiation field in vacuum has been constructed in the cylinder case, where state variables have been defined globally. It was impossible in the cavity case to define state variables globally, since the system is anisotropic and inhomogeneous. However, after introducing the idea of local steady states which is motivated by the anisotropy and inhomogeneity of the system, we have been able to extend the SST to the cavity case. 
As mentioned in the second paragraph in subsection \ref{sec-cylinder.distribution}, the state variables in the SST for a radiation field are given by an appropriate linear combination of two equilibrium values.

It should be emphasised that, contrary to the cases of ordinary dissipative systems \cite{ref-cit} \cite{ref-eit} \cite{ref-s.t}, the energy flux can not work as a consistent state variable for a radiation field \cite{ref-essex} \cite{ref-wildt}, as has been explicitly shown in the equation (\ref{eq-cylinder.j.conjugate}). Concerning this issue, we have found in this paper that the consistent state variables of the nonequilibrium order of steady states are the difference of the temperatures of two black bodies $\tau$ and its conjugate variable $\psirad{x}$. A somewhat detailed discussion of the energy flux is put in appendix \ref{app-flux}, where the reason why an energy flux can not behave as a state variable is explored.

{\bf Special properties of a radiation field}: Other than the exclusion of the energy flux from consistent state variables, there are other special properties of a radiation field. As mentioned below the first law (\ref{eq-cylinder.1st}), the constant volume strain has no physical meaning in the first law for a radiation field. Further the first law (\ref{eq-cylinder.1st}) together with the Legendre transformation (\ref{eq-cylinder.legendre}) gives,
\sikibnon
 dF_{rad} = - S_{rad} \, dT_{rad} - P_{rad} \, dV - \Psi \, d\tau \, .
\sikienon
This shows that the components of the SST pressure tensor $\pt$ can not be obtained from the SST free energy $F_{rad}$. Such a limitation on the information obtained from the free energy should be understood as a unique property of a radiation field. 

However, if one proceeds to a general relativistic extension, it may be possible to obtain the components of $\pt$ from the free energy. The pressure tensor should be extended to the so-called {\it stress-energy-momentum tensor} ${\bf T}$ in the context of general relativity. The tensor ${\bf T}$ is related to the deformation ${\bf R}$ ({\it curvature tensor} of rank two) of the spacetime via the Einstein equation, and the tensor ${\bf R}$ can be interpreted as the strain of the spacetime raised by the radiation field. That is, the work term may be given as ${\bf T:\,} d{\bf R}$. Hence we may obtain ${\bf T}$ by the partial derivative of the free energy by ${\bf R}$. 

{\bf The distance of steady states from equilibrium states}: 
Let us remark on the method of finding the equilibrium limit and on the definition of the intensive nonequilibrium order parameter $\tau$. In constructing the SST for the cavity case, we have considered an equilibrium limit as $T_{in} - T_{out} \to 0$. However there is another method of finding the equilibrium limit which is to remove the inner black body. This limit is taken with 
$\gin{x} \to 0$ ($\Leftrightarrow \gout{x} \to 1$). With this limit, $\tau$ remains at a constant value $T_{in} - T_{out}$ before $\gin{x}$ reaches zero, but $\tau$ becomes $T_{out}$ discontinuously at $\gin{x} = 0$. That is, $\tau$ does not disappear at this equilibrium limit, and not satisfy the requirement (R2). However, because $\psirad{x} \to 0$ for both equilibrium limits $T_{in} - T_{out} \to 0$ and $\gin{x} \to 0$, the violation of the requirement (R2) by $\tau$ is harmless and our formulation of the SST remains consistent.

If one wishes to modify the definition of $\tau$ to satisfy the limit $\tau \to 0$ as $\gin{x} \to 0$, one method is to re-define as $\tau = ( g_{in} g_{out} )^q \, ( T_{in} - T_{out})$, where $q$ is a constant and $0 < q < 1$. Consequently $\psirad{x}$ should be modified to $\psirad{x} = ( g_{in} g_{out} )^{1-q} \, ( s_{eq}(T_{in}) - s_{eq}(T_{out}) )$. However as discussed in the previous paragraph, $q = 0$ does not cause inconsistency in formulating the SST. Therefore we adopt $q = 0$ as the simplest definition of $\tau$. 

The nonequilibrium order parameters $\tau$ and $\psirad{x}$ seem to give some measure of distance of steady states from equilibrium states. It is appropriate to require that the distance of steady states from equilibrium states becomes zero with both equilibrium limits, $T_{in} - T_{out} \to 0$ and $\gin{x} \to 0$. Note that, when we adopt the simplest definition of $\tau \equiv T_{in} - T_{out}$, this $\tau$ can not disappear with the limit $\gin{x} \to 0$. Therefore it is plausible to consider $\psirad{x}$ as the measure of distance of steady states from equilibrium states, where this distance should be interpreted as a kind of averaged distance from two equilibrium states of temperatures $T_{in}$ and $T_{out}$. The larger the value of $\psirad{x}$, the more distant a steady state from two equilibrium states. In fixing the values of $T_{in}$ and $T_{out}$, the distance of a steady state from two equilibrium states can vary by changing the geometry of system. Here note an inequality, $0 < \alpha \, (1 - \alpha) \leq 1/4$ for $0 < \alpha < 1$, where the maximum value $1/4$ is taken with $\alpha = 1/2$. Then we find that $\psirad{x}$ takes the largest value for the case $\gin{x} = \gout{x} = 1/2$. On the other hand, the cylinder case is recovered from the cavity case by setting $\gin{x} = \gout{x} = 1/2$. Consequently, it is concluded that the radiation field in the cylinder case is in the most distant steady state from two equilibrium states. This implies that the higher the geometrical symmetry of the system, the more distant the steady state from two equilibrium states.

{\bf Future improvements}: 
We have assumed the condition (C1). In a practical experiment, the condition (C1) may imply that the time scales of thermalisation of the two black bodies are very short. Then a heat flux probably penetrates well into the black bodies, and their surfaces are affected by the heat flux. In such a case, the distribution functions (\ref{eq-cylinder.distribution}) and (\ref{eq-cavity.distribution}) should be extended to nonequilibrium ones. Therefore, because the radiation field is completely characterised by the temperatures at the surfaces of the two bodies, the nonequilibrium effect of the surfaces probably becomes essential for the SST for a radiation field.

If we consider $T_{out} = 0$ for the cavity case, then no energy flux comes into the inner black body and no nonequilibrium effect on its surface arises. Such a case can be considered in a general relativistic system. It has already been known that a {\it black hole} emits a thermal radiation, that is, a black hole can be considered as a black body \cite{ref-bh-1} \cite{ref-bh-2}. Then, if a black hole is isolated from other celestial bodies, no energy flux may come into the black hole. Therefore the cavity case with $T_{out} = 0$ can be applied to the system composed of a black hole and its thermal radiation. However before proceeding to such an application, we may need to carry out a relativistic extension on the SST for a radiation field. 

{\bf Applications}: 
In laboratory experiments which need to keep an instrument isolated, a method used frequently for heat insulation is to keep the instrument be surrounded by vacuum, where the instrument corresponds to the inner black body in the cavity case. A total value of the nonequilibrium order of a radiation field in the surrounding vacuum is estimated as $\Psi = \int_{V_{rad}} dx^3 \psirad{x} \sim ( T_{in}^{\,3} - T_{out}^{\,3} ) \, V_{rad} \times 10^{-8}$ erg/K, where we set $g_{in} \sim g_{out} \sim O(1)$. For example, consider the case that the instrument is sufficiently near an equilibrium state whose equation of state is of an ideal gas type. Then the entropy of the instrument is estimated as $S_{in} \sim ( E_{in} + P_{in} V_{in} )/T_{in} \sim n \, N_A \, k_B \sim n \times 10^7$ erg/K, where $n$ is a numerical factor which denotes the size of the instrument and $N_A$ is the Avogadro's number. Then we find, $\Psi/S_{in} \sim ( T_{in}^{\,3} - T_{out}^{\,3} ) \, V_{rad}/n \times 10^{-15}$. This denotes that an effect of a radiation field in the surrounding vacuum does not come into account unless we have sufficiently large difference of temperatures, $\tau \gtrsim 10^5$ K, or sufficiently large volume of the surrounding vacuum, $V_{rad} \gtrsim 10^{15}$ cm$^3$, or sufficiently small instrument, $n \lesssim 10^{-15}$. That is, the heat insulation with surrounding an instrument by vacuum is very effective in laboratory experiments. 

A situation in which an effect of the SST for a radiation field becomes important can be considered in astrophysics and cosmology, since the volume of vacuum space is sufficiently large in inter-stellar spaces and temperatures of celestial bodies can be sufficiently large. When a radiation field in an inter-stellar space is considered, the condition (C2) is violated and we need to take the effects of the retarded time on photons into account. However, when a radiation field inside a star is considered, it is possible to obtain a notable implication for astrophysics with the present form of the SST for a radiation field. A somewhat detail explanation of this astrophysical implication is summarised in appendix \ref{app-implication}. The point of this implication is that, in comparison with the usual analysis of a self-gravitating gas system using the ordinary (local) equilibrium thermodynamics, the SST for a radiation field tends to prevent the so-called {\it gravothermal catastrophe} from occurring. The gravothermal catastrophe is one of the most significant phenomena of self-gravitating systems, which is the essential physical basis to characterise the formation and evolution of astrophysical objects like stars and galaxies \cite{ref-stellar} \cite{ref-galactic}. Therefore the SST for a radiation field may give some useful corrections or developments in the theories of stellar structure and other astrophysical systems.

\begin{acknowledgments}
I would like to express my gratitude to Hiroshi Harashina, Kenji Imai, Hiromi Kase and Shin-ichi Sasa for their useful comments in early stage of my study on the subject discussed in this paper. And I would also like to thank referee of this paper for his/her questions which led me to refine some parts of this paper. 
\end{acknowledgments}

\appendix

\section{Energy flux}
\label{app-flux}

Let us consider the cavity case. Look at a spatial point $\vec{x}$ in the radiation field, and take the sum $\vec{p}_{tot}$ of the momentums $\vec{p}_{in}$ and $\vec{p}_{out}$ of all photons passing the point $\vec{x}$ at one moment. The direction of a net energy flow through $\vec{x}$ at that moment is parallel to $\vec{p}_{tot}$. Then we can define the energy flux vector as
\sikibnon
 \vec{j} \equiv j \, \vec{n} \, ,
\sikienon
where $\vec{n}$ is the unit vector parallel to $\vec{p}_{tot}$ at $\vec{x}$, and the magnitude of this vector $j$ is the energy flux given by
\sikibnon
 j \equiv 2 \int \frac{dp^3}{h^3} \, \hbar \, \omega \, d(\vec{x},\vec{p}) \, c \, \cos\phi \, ,
\sikienon
where the factor $2$ is due to the helicity of a photon and $\phi$ is the angle between $\vec{p}$ and $\vec{n}$ given by $p \cos\phi = \vec{n} \cdot \vec{p}$. With the relation $\int_0^{\infty} dx x^3 / (e^x -1) = \pi^4 / 15$, we obtain
\sikibnon
 j(\vec{x}) =
  \sigma \, \left(\, \gamma_{in}(\vec{x}) \, T_{in}^{\,4}
   + \gamma_{out}(\vec{x}) \, T_{out}^{\,4} \, \right) \, ,
\sikienon
where $\gamma_{in}(\vec{x})$ and $\gamma_{out}(\vec{x})$ are the geometrical factors given as
\sikibnon
 \gamma_{in}(\vec{x})
  = \frac{1}{\pi} \int_{\vec{p}_{in}} d\Omega_{xp} \, \cos\phi \quad , \quad
 \gamma_{out}(\vec{x})
  = \frac{1}{\pi} \int_{\vec{p}_{out}} d\Omega_{xp} \, \cos\phi \, ,
\sikienon
where $d\Omega_{xp}$ is the solid-angle element in $\vec{p}$-space at $\vec{x}$. These satisfy the relation, $\gamma_{in}(\vec{x}) + \gamma_{out}(\vec{x}) = 0$, then $\vec{j} = 0$ at the equilibrium limit $T_{in} = T_{out}$. The factors $\gamma_{in}(\vec{x})$ and $\gamma_{out}(\vec{x})$ are obviously different from the factors $\gin{x}$ and $\gout{x}$ which appear in the form of the state variables given in section \ref{sec-cavity}. This difference may be interpreted as one reason why the energy flux does not work as a consistent state variable.

\section{Implication for astrophysics}
\label{app-implication}

\begin{figure}[t]
 \begin{center}
 \includegraphics[height=25mm]{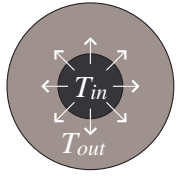}
 \end{center}
\caption{A toy model for the gravothermal catastrophe. The heat capacity of the inner black body is negative.}
\label{pic.6}
\end{figure}

A remarkable property of a self-gravitating system is the negative heat capacity of a region around the centre of mass due to the virial theorem and the absence of the lower bound of the gravitational potential ($\propto -1/r$, where $r$ is the distance from the centre of mass) \cite{ref-stellar} \cite{ref-galactic}. The significant phenomenon caused by the negative heat capacity is the {\it gravothermal catastrophe}, which is, for example, the essential basis to cause and maintain the nuclear reactions in the core of the star. The simplest toy model of the gravothermal catastrophe is given by setting the volume of the radiation field in our cavity case as zero (see figure \ref{pic.6}), and setting the heat capacity of the inner black body as negative, $C_{in} < 0$. Let us consider the case $T_{in} > T_{out}$. Then, while the temperature of the outer black body continues to increase due to the incoming heat from the inner black body, the temperature of the inner black body does also continue to increase. Further if the relation $\left| C_{in} \right| < C_{out}$ holds, the rate of temperature increase of the inner black body is faster than that of the outer black body. Then the difference of the temperatures continues to increase without upper limit. This is a rough illustration of the gravothermal catastrophe. For an example of the case of a star formation, the inner black body corresponds to the core of the star and the outer black body to the other part of the star, and the temperature increase of the core due to the gravothermal catastrophe triggers the nuclear interactions in the core. Here note that, since the virial theorem is used for deriving the negative heat capacity, the gravothermal catastrophe is a consequence of a local equilibrium treatment of the system under consideration. And, even if the radiative energy transfer is also included in the above discussion, the gravothermal catastrophe should be concluded whenever the radiative energy transfer is treated in a traditional manner mentioned as in the beginning of section \ref{sec-intro}. 

Let us again place a radiation field between the inner and outer black bodies, and examine the relaxation process with the negative heat capacity of the inner black body $C_{in} < 0$. The same calculation given in subsection \ref{sec-cavity.2nd} holds in this case. Here look at the total energy and its differential,
\sikibnon
 E_{tot} &=&
    \left(\, E_{in} + \frac{4\,\sigma}{c}\, G_{in} \, T_{in}^{\,4} \,\right)
  + \left(\, E_{out} + \frac{4\,\sigma}{c}\, G_{out} \, T_{out}^{\,4} \,\right) \, , \\
 dE_{tot} &=&
    \left(\, C_{in} + \frac{16 \, \sigma}{c} \, G_{in} \, T_{in}^{\,3} \,\right) \, dT_{in}
  + \left(\, C_{out}
           + \frac{16 \, \sigma}{c} \, G_{out} \, T_{out}^{\,3} \,\right) \, dT_{out} \, ,
\sikienon
where the terms including $G_{in}$ or $G_{out}$ come from the SST internal energy $E_{rad}$. This enables us to consider our system as being composed of two subsystems X and Y whose heat capacities $C_X$ and $C_Y$ are given as
\sikibnon
 C_X = C_{in} + \frac{16 \, \sigma}{c} \, G_{in} \, T_{in}^{\,3} \quad , \quad
 C_Y = C_{out} + \frac{16 \, \sigma}{c} \, G_{out} \, T_{out}^{\,3} \, ,
\sikienon
and the temperatures of the subsystems are $T_{in}$ for X and $T_{out}$ for Y. For simplicity set $C_{in} =$ constant $< 0$ and $T_{in} > T_{out}$. In this case, energy flows from the subsystem X to Y, and the heat capacity $C_X$ of X is negative for a sufficiently low temperature of the inner black body $T_{in}$. Therefore $T_{in}$ increases. However, after $T_{in}$ increases sufficiently, the heat capacity $C_X$ turns from negative to positive. This denotes that the evolution of $T_{in}$ during the relaxation stabilises at a critical temperature $T_{in}^{\, \ast}$ determined by $C_X = 0$. Hence it is expected that the whole system settles in the equilibrium state of the critical temperature $T_{in}^{\,\ast}$. This discussion is applicable to the case $\left| C_{in} \right| < C_{out}$, which results in the gravothermal catastrophe when the system under consideration is treated with a local equilibrium theory. Therefore it is implied that the SST for a radiation field tends to prevent the gravothermal catastrophe from occurring. The theory of the evolution of the stellar structure and the rate of star-formation in a galaxy may be corrected if the SST for a radiation field will be applied to the radiation field inside a star. 

\end{document}